# PM²PLS: An Integration of Proxy Mobile IPv6 and MPLS

Carlos A. Astudillo[1], Oscar J. Calderón[1] and Jesús H. Ortiz[2]

[1] New Technologies in Telecommunications R&D Group, Department of Telecommunications, University of Cauca, Popayán, 19003, Colombia

[2] School of Computer Engineering, University of Castilla y la Mancha
Ciudad Real, Spain

**Abstract**
This paper proposes a handover scheme supporting Multi-Protocol Label Switching (MPLS) in a Proxy Mobile IPv6 (PMIPv6) domain that improves the mobility and gives Quality of Service (QoS) and Traffic Engineering (TE) capabilities in wireless access networks. The proposed scheme takes advantages of both PMIPv6 and MPLS. PMIPv6 was designed to provide NETwork-based Localized Mobility Management (NETLMM) support to a Mobile Node (MN); therefore, the MN does not perform any mobility related signaling, while MPLS is used as an alternative tunneling technology between the Mobile Access Gateway (MAG) and the Local Mobility Anchor (LMA) replacing the IP-in-IP tunnels with Label Switched Path (LSP) tunnels. It can also be integrated with other QoS architectures such as Differentiated Services (DiffServ) and/or Integrated Services (IntServ). In this study, we used MATLAB to perform an analysis to evaluate the impact of introducing MPLS technology in PMIPv6 domain based on handover latency, operational overhead and packet loss during the handover. This was compared with PMIPv6, and a PMIPv6/MPLS integration. We proved that the proposed scheme can give better performance than other schemes.
*Keywords: Localized Mobility Management, MPLS, PMIPv6, PMIPv6/MPLS, PM²PLS.*

## 1. Introduction

Some host-based mobility management protocols such as Mobile IPv6 (MIPv6) [1] and its extensions (i.e. Hierarchical Mobile IPv6 (HMIPv6) [2] and Fast Handover in Mobile IPv6 (FMIPv6) [3]) have been standardized by the Internet Engineering Task Force (IETF) for Internet mobility support, but they have not widely deployed in real implementations [4]. One of the most important obstacles in order to deploy mobility protocols is the modification that must be done in the terminal (Mobile Host - MH). Proxy Mobile IPv6 has been proposed by the IETF NETLMM working group as a network-based mobility management protocol [5]. It allows the communication between the Mobile Node and the Correspondent Node (CN) while MN moves without its participation in any mobility signaling. On the other hand, Multiprotocol Label Switching is a forwarding technology that supports Quality of Service and Traffic Engineering capabilities in IP networks [6]. Furthermore, it provides fast and efficient forwarding by using labels swapping instead of IP forwarding.

MPLS is being used by most network operators to carry IP traffic. Introduce network-based mobility capabilities in MPLS networks can be useful [7].

There are few works that have handled the integration of PMIPv6 and MPLS. Recently, an IETF Internet Draft proposed MPLS tunnels (LSP tunnels) as an alternative to IP-in-IP tunnel between Local Mobility Anchor (LMA) and Mobile Access Gateway (MAG) [7]. The draft specifies two different labels: a classic MPLS label and Virtual Pipe (VP) labels as a way to differentiate traffic in the same tunnel. The authors focus on the management of VP labels rather than classic MPLS labels. The authors assume that there are LSPs established between the MAG and the LMA and use two labels for each packet; both labels are pushed by the Label Edge Router (LER).

But, as mentioned in [8], the use of VP label is not strictly necessary because this label is only used to eliminate the necessity of the LMA to look up the network layer header in order to send packets to the CN. It adds 4 overhead bytes (VP label size) to the LSP tunnel (8 overhead bytes in total). Reference [8] makes a study of PMIPv6/MPLS on Wireless Mesh Network (WMN) with and without VP labels in terms of handover delay and operation overhead. Reference [9] makes a study in an Aeronautical Telecommunication Network (ATN) and uses VP labels in the same way of [7]. Reference [10] makes a quantitative and qualitative analysis of the PMIP/MPLS integration and other schemes, but they do not give details about design considerations, label management or architecture operation.

This work proposes an integration of PMIPv6 and MPLS called PM²PLS. The integration is done in an overlay way [11] and the relationship between binding updates and LSPs setup is sequential. We do not consider necessary to use VP label since this label only divided traffic from different operators (its use is optional). We use Resource Reservation Protocol – Traffic Engineering (RSVP-TE) [12] as label distribution protocol to establish a "bidirectional LSP" between the LMA and the MAG. Since a LSP in MPLS is unidirectional, we call "bidirectional LSP" to two LSP that do not necessarily follow the same upstream and downstream path but that the ingress Label Switch Router (LSR) in the LSP upstream is the egress LSR in the LSP downstream and vice verse. In future works, we want to integrate PM$^2$PLS and QoS architectures such as IntServ and/or DiffServ in order to assure QoS in a mobility enabled MPLS access network where the MN is not based on MIPv6.

The rest of the paper is organized as follows. Section 2 presents an overview about PMIPv6 and MPLS. Section 3 introduces the PMIPv6/MPLS integration called PM$^2$PLS. Section 4 shows the performance analysis of PM$^2$PLS on 802.11 access network based on handover latency, operational overhead and packet loss during handover. Finally, we conclude in Section 5.

## 2. Background

2.1 Proxy Mobile IPv6

PMIPv6 was designed to provide network-based mobility support to a MN in a topologically localized domain [5]; this means that the CN is exempted to participate in any mobility related signaling and all mobility control functions shift to the network. In this context, PMIPv6 defined two new entities called Local Mobility Anchor and Mobile Access Gateway. The function of LMA is to maintain reachability to the MN and it is the topological anchor point for the MN's home network prefix(es), this entity has a Binding Cache (BC) that links the MN with its current Proxy CoA (MAG's address). MAG runs in the Access Router (AR) and is responsible for tracking the mobile node´s movements at the access link and for initiating binding registrations to the LMA; it also establishes a bidirectional tunnel with the LMA to enable the MN to use an address from its home network prefix (MN-HNP) and emulates the MN's home link. This entity has a Binding Update List (BUL) which contains the MNs attached to it, and their corresponding LMAA (LMA's address). Figure 1 shows a common PMIPv6 scenario with LMAs, MAGs, MNs, CN, tunnels between LMA and MAG and data flow.

In a PMIPv6 domain, the options for establishing the tunnel between LMA and MAG are as follows: IPv6-In-IPv6 [5], Generic Routing Encapsulation (GRE), IPv6-In-IPv4 or IPv4-In-IPv4 [13].

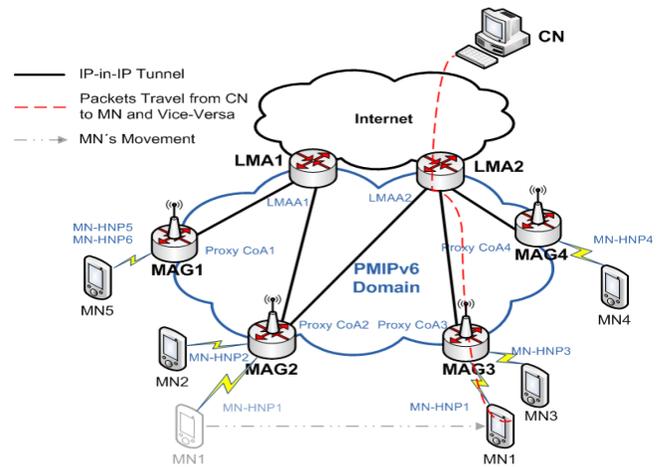

Fig. 1    PMIPv6 scenario.

2.1 Multi-Protocol Label Switching

Conventional IP forwarding mechanisms are based on network reachability information. As a packet traverses the network, each router uses the IP header in the packet to obtain the forwarding information. This process is repeated at each router in the path, so the optimal forwarding is calculated again and again. MPLS [6] is a forwarding packets paradigm integrated with network-layer routing. It is based on labels that assign packet flows to a Forwarding Equivalent Class (FEC). FEC has all information about the packet (e.g. destination, precedence, Virtual Private Network (VPN) membership, QoS information, route of the packet, etc.), once a packet is assigned to a FEC no further analysis is done by subsequent routers, all forwarding is driven by the labels. All packets with the same FEC use the same virtual circuit called Label Switched Path (LSP). To deploy MPLS in an IP network, a label header is inserted between layer two and layer three headers as shown in Figure 2. The MPLS header is composed by: 20-bit label field, 3-bit initially defined as EXPerimental and current used as Traffic Class (TC) field [15], 1-bit Bottom of Stack (S) field, and 8-bit Time to Live (TTL) field. MPLS also offers a traffic engineering capabilities that provides better use of the network resources.

　MPLS consists of two fundamentals components: The FEC-to-NHLFE mapping (FTN) which forwards unlabeled packets, this function is running in the ingress router (LER, Label Edge Router) and mapping between IP packets and FEC must be performed by the LER. And the Incoming Label Mapping (ILM) that makes a Label-to-NHLFE mapping to forward labeled packets.
The RFC 3031 defines a "LSP Tunnel" as follows:  "It is possible to implement a tunnel as a LSP, and use label switching rather than network layer encapsulation to cause the packet to travel through the tunnel" [6]. The packets that are sent through the LSP tunnel constitute a FEC.

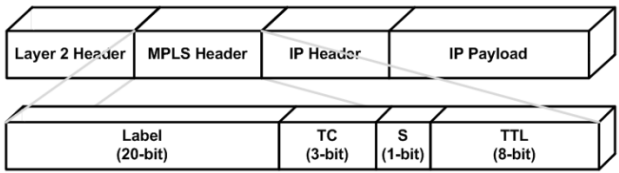

Fig. 2   MPLS header format.

## 3. PMIPv6 and MPLS Integration

We propose a PMIPv6/MPLS architecture called PM²PLS. First, we give previous concepts on the integration of MPLS and MIPv6 (and its extensions), then, we describe the design considerations, MAG and LMA operation and finally, the signaling flow between components is described.

3.1 Previous Concepts

Previous works on integrating MIPv6, HMIPv6 and/or FMIPv6 in MPLS networks consider two models for doing that: integrated or overlay [11]. In the integrated model, some processes are united; in the overlay one, processes and information are separated as long as possible. We choose to use the overlay model since it allows an easy integration with current deployed MPLS networks.

Another important item in previous integrations is the relationship between binding updates and LSPs setup. There are two proposes. The first one is to make the LSP setup in an encapsulated way [11] which means that the LSP establishment is initialized after a Binding Update (BU) message arrives to the Home Agent (HA), Mobility Anchor Point (MAP) or Regional Gateway (RG) but the Binding Acknowledgment (BA) is sent after a LSP setup process is finished. The other method is called "sequential" where the LSP setup is initialized after a successful binding update process finished [11]. It means that the LSP setup is initialized when a BA message arrives to CN, Foreign Agent (FA) or Access Router (AR). Reference [11] concluded that sequential way has better handover performance than encapsulated one. In our scheme the relationship between binding updates and LSP setup can be viewed as "sequential", but we optimized the LSP setup since the process is initialized in the LMA after the Proxy Binding Update (PBU) message has been accepted and Proxy Binding Acknowledgment (PBA) message sent, it does not wait for PBA arrives to the MAG since we consider that it is not necessary.

3.2 Design Considerations

We give the design considerations for the PM²PLS architecture in this subsection.

- We used LSP tunnels as specified in [6], [12]. The LSP Tunnel must be "bidirectional" between MAG and LMA (two LSP Tunnels established by RSVP-TE, one from LMA to MAG and other between MAG and LMA). Note that the upstream LSP not necessarily follows the same path that downstream LSP. This "bidirectional" LSP Tunnel must be used for forwarding the mobile nodes' data traffic between MAG and LMA. It can also be used for sending PBU and PBA between MAG and LMA.
- The LSP setup could be pre-established or dynamically assigned. In a dynamic way, the LSP would be setup only once, when the first MN arrives to specific MAG, the follows MNs can used the established LSP, if it is necessary to re-evaluated the LSP capabilities, it should be performed by RSVP-TE techniques. It also improves the Proxy Binding Update and Proxy Binding Acknowledgment messages delivery of sub-sequence location updates.
- The introduction of network-based mobility in MPLS networks should be in an overlay way. It means that data base will not be integrated between PMIPv6 and MPLS. The BC, BUL and the Label Forwarding Information Base (LFIB) should be maintained separately. But a relationship between processes sequence should be performed and the information should be shared.
- The MN should be IPv6-Base. We only consider the use of IPv6 MN-HoA since the process of address configuration in IPv4 is too large, instead IPv6 supports stateless address configuration.
- The Transport Network could be IPv6 or IPv4.
- The traffic in the same MAG is managed for itself.
- The wireless access network that we consider in this study is 802.11. It is necessary to define the Access Network (AN) type because of the analysis that will be described, but it does not imply that others access technologies as Long Term Evolution (LTE), WiMax or 3G Networks couldn't be used with PM2PLS.
- This architecture cannot support multicast traffic.
- Penultimate hop popping is desirable. It should be used, since the packet processing at the last hop (in the MPLS domain) would be optimized. It avoids double processing in the last hop (i.e. MPLS and IP header processing).
- Label merging and aggregation are undesirable. Those constraints allow having unique label per LSP and more than one LSP for the same FEC, respectively (e.g. it is useful when we want to introduce load balancing between the LMA and a specific MAG).

## 3.3 Architecture Components

The architecture components shown in Figure 3 are described. Figure 4 gives the protocol stack of PM²PLS entities and the signaling flow between them when a handover occurs is shown in Figure 5.

- MAG/LER: It is an entity which has the MAG (from PMIPv6) and LER (from MPLS) functionality inside its protocol stack.
- LMA/LER: It is an entity which has the LMA (from PMIPv6) and LER (from MPLS) functionality inside its protocol stack.
- LSR: It is a MPLS router as specified in [6].
- MN: It is a mobile node which implements IPv6.
- CN: It is a mobile/fixed node which implements IPv6 or IPv4.

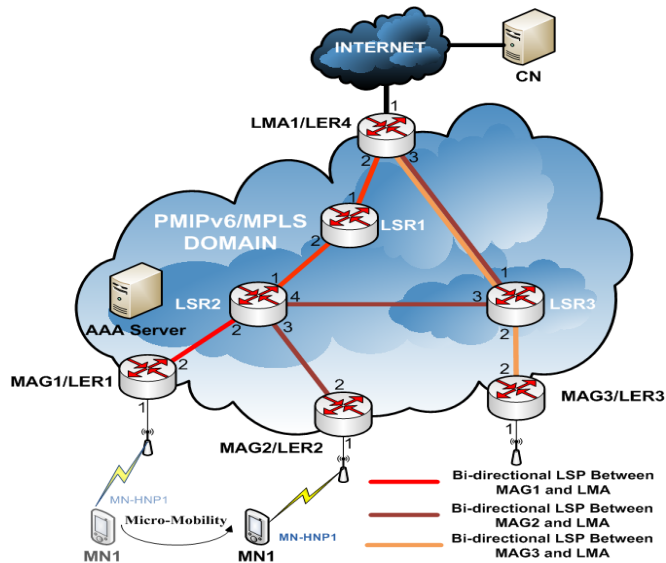

Fig. 3   PM²PLS scenario.

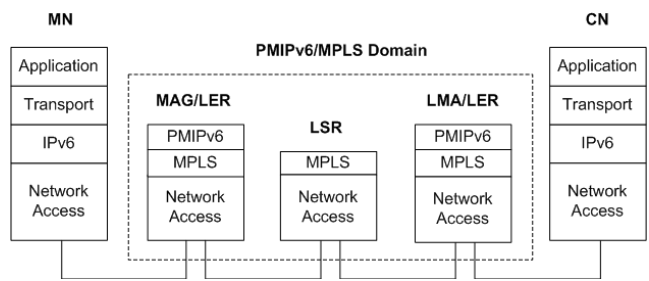

Fig. 4   Protocols stack of PM²PLS components

## 3.4 LMA/LER Operation

When a PBU message is received by the LMA, it processes the message as specified in [5], after PBU is accepted and the PBA is sent, immediately the LMA verifies if it is assigned the MN's PCoA to a FEC (there are LSP tunnel between LMA and MN's MAG). If an entry already exists with the MN-PCoA as FEC, it does not need to setup the LSP, since a LSP Tunnel already exists, If not a RSVP Path message are generated from LMA to MAG to setup the LSP between LMA and MAG. When the LSP setup process is finished (Path and Resv RSVP messages are received and processed) and the LMA had assigned a label to that FEC, it should have a entry in the LFIB with the FEC assign to the tunnel between LMA and MAG. Periodically, the LSP capability should be evaluated in order to assure that the traffic across the LSP is being satisfied.

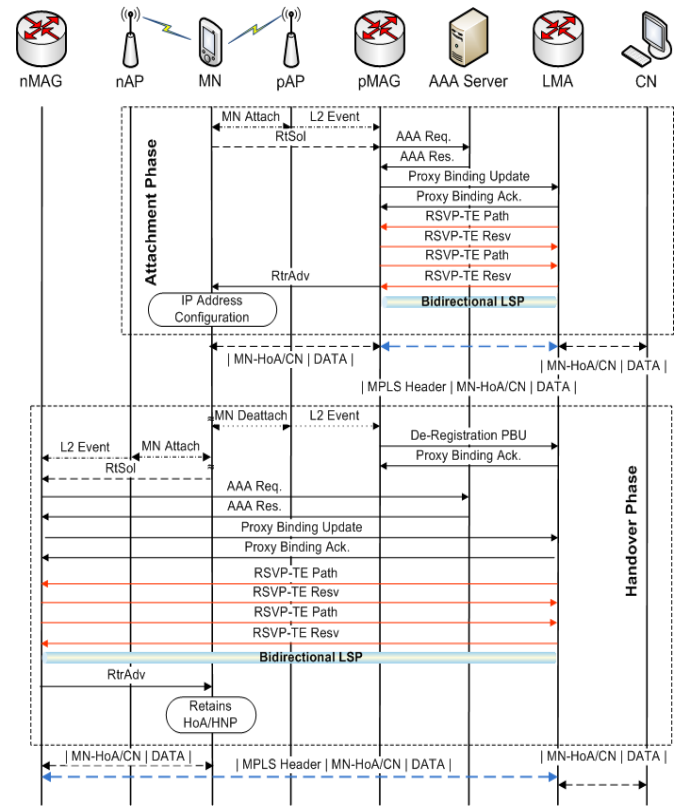

Fig. 5   Signaling flow in PM²PLS.

## 3.5 MAG/LER Operation

When a PBA message is received by the MAG with a status field set to zero (accepted), it processes the message in the same way as specified in [5], and then a RSVP Path message is generated from MAG to LMA to setup the LSP between MAG and LMA. If an entry already exists with MN´s LMA as a FEC, it does not need to setup the LSP, since it already exists. Periodically, the LSP capability should be evaluated in order to assure that the traffic across the LSP is being satisfied.

### 3.6 Handover Procedure

When roaming for first time in a PMIPv6/MPLS domain, the MN obtains a MN-HoA based on its HNP and keeps it as long as stays in the PMIPv6 domain. This means that the MN only executes the address configuration and Duplicate Address Detection (DAD) once.

The handover process in PM$^2$PLS scenario is as follows. When the MN moves from a MAG/LER to another MAG/LER in the same domain, first the MN detaches from a Access Point (AP) in a previous MAG/LER (pMAG/LER) area and attaches to a AP in new MAG/LER (nMAG/LER) area, at this moment nMAG/LER knows the MN-ID and other information by layer 2 procedures (Note that in PMIPv6 it is not necessary to wait for a Router Solicitation message (RtSol), this message can be sent by the MN at any time during the handover process). nMAG/LER performs a MN's authentication, and then sends a PBU to the LMA. Upon receiving the PBU message, the LMA follows the procedure described in section 3.4, it generates a PBA messages and if it is necessary to send RSVP Path message. The MAG on receiving the PBA message follows the procedure described in section 3.5. It updates its Binding Update List and sends a RSVP-Path if it is necessary. Finally, the sends a Router Advertisement (RtrAdv) message containing the MN's HNP, and this will ensure the MN will not detect any change with respect to the layer 3 attachment of its interface (it retains the configured address).

### 3.7 Example of LFIBs in PM$^2$PLS Nodes

Based on Figure 3, we give an example of the Label Forwarding Information Base (LFIB) of each node in the PM$^2$PLS scenario. In this example, we use penultimate hop popping and assume that the upstream LSP has the same path (the same nodes) of the downstream LSP. We show the content of the LFIB in LMA1/LER4 (Table 1), MAG1/LER1 (Table 2), MAG2/LER2 (Table 3), MAG3/LER3 (Table 4), LSR1 (Table 5), LSR2 (Table 6), and LSR3 (Table 7).

## 4. Performance Analysis

In this section we analyze the performance of PM$^2$PLS on 802.11 Wireless LAN (WLAN) access network based on handover delay, attachment delay, operational overhead and packet loss during handover. We compared our proposal with single PMIPv6 and PMIPv6/MPLS in an encapsulated way as proposed in [8].

Table 1: LMA1/LER4's LFIB

| FEC | In Label | In IF | Out Label | Out IF |
|---|---|---|---|---|
| LMA-MAG1 | - | - | 20 | 2 |
| LMA-MAG2 | - | - | 22 | 3 |
| LMA-MAG3 | - | - | 27 | 3 |

Table 2: MAG1/LER1's LFIB

| FEC | In Label | In IF | Out Label | Out IF |
|---|---|---|---|---|
| MAG1-LMA | - | - | 40 | 2 |

Table 3: MAG2/LER2's LFIB

| FEC | In Label | In IF | Out Label | Out IF |
|---|---|---|---|---|
| MAG2-LMA | - | - | 55 | 2 |

Table 4: MAG3/LER3's LFIB

| FEC | In Label | In IF | Out Label | Out IF |
|---|---|---|---|---|
| MAG3-LMA | - | - | 60 | 2 |

Table 5: LSR1's LFIB

| FEC | In Label | In IF | Out Label | Out IF |
|---|---|---|---|---|
| LMA-MAG1 | 20 | 1 | 15 | 2 |
| MAG1-LMA | 35 | 2 | - | 1 |

Table 6: LSR2's LFIB

| FEC | In Label | In IF | Out Label | Out IF |
|---|---|---|---|---|
| LMA-MAG1 | 15 | 1 | - | 2 |
| MAG1-LMA | 40 | 2 | 35 | 1 |
| LMA-MAG2 | 32 | 4 | - | 3 |
| MAG2-LMA | 55 | 3 | 50 | 4 |

Table 7: LSR3's LFIB

| FEC | In Label | In IF | Out Label | Out IF |
|---|---|---|---|---|
| LMA-MAG2 | 22 | 1 | 32 | 3 |
| MAG2-LMA | 50 | 3 | - | 1 |
| LMA-MAG3 | 27 | 1 | - | 2 |
| MAG3-LMA | 60 | 2 | - | 1 |

### 4.1 Handover Process in 802.11

In order to study the handover performance of PM$^2$PLS, we consider an 802.11 WLAN access to calculate the L2 handover delay (that is when a MN attaches to a new Access Point (AP)). During the handover at layer two, the station cannot communicate with its current AP. The IEEE 802.11 handover procedure involves at least three entities: the Station (MN in PM$^2$PLS), the Old AP and the New AP. It is executed in three phases: Scanning (Active or Passive), Authentication and Re-association as shown in Figure 6 [16]. The scanning phase in a handover process is attributed to mobility, when signal strength and the signal-to-noise ratio are degraded the handover starts. At this point, the client cannot communicate with its current AP and it initializes the scanning phase. There are two methods in this phase: Active and Passive. In the passive method the station only waits to hear periodic beacons transmitted by neighbour APs in the new channel, in the active one, the station also sends probe message on each channel in its list and receives response of APs in its coverage range. When the station finds a new AP, it sends an authentication message, and once authenticated can send the re-association message. In this last phase includes the IAPP (Inter Access Point Protocol) [17] procedure to transfer context between Old AP and New AP.

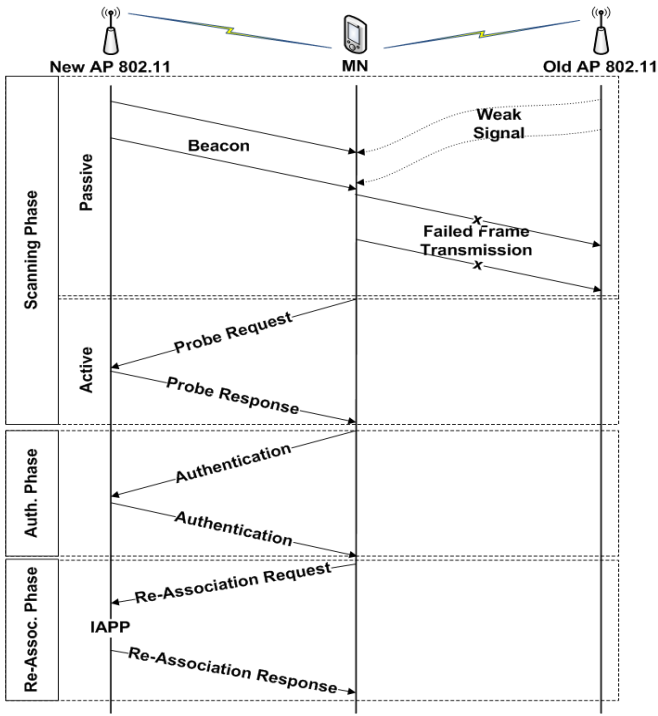

Fig. 6   802.11 handover process

Table 8: Parameter descriptions/settings

| Parameter | Description | Value |
|---|---|---|
| $\alpha_{RP}$ | IP router processing time. | 0.2 ms |
| $\alpha_{AAA\text{-}Server}$ | Processing time of AAA Server. | 0.1 ms |
| $t_{x,y}$ | Time required for a message to pass through links from node x to node y | N/A |
| $t_{WL}$ | Wireless link delay. | 10 ms [4] |
| $t_{Scanning}$ | Delay due to scanning phase of 802.11. | 100 ms [16] |
| $T_{REG}$ | Registration or binding update delay. | N/A |
| $t_{PBU}$ | Time of Proxy Binding Update message | N/A |
| $t_{PBA}$ | Time of Proxy Binding Acknowledgment message | N/A |
| $T_{MD}$ | Mobility detection delay. | 0 ms |
| $T_{L3HO}$ | L3 handover delay. | N/A |
| $T_{L2HO}$ | L2 handover delay. | 115 ms [4] |
| $T_{HO}$ | Total handover delay. | N/A |
| $T_{Bi\text{-}LSP\text{-}Setup}$ | Delay due to bidirectional LSP setup. | N/A |
| $t_{Autentication}$ | Delay due to 802.11 authentication phase. | 5 ms [16] |
| $t_{Assocciation}$ | Delay due to 802.11 association phase. | 10 ms [16] |
| $t_{AP\text{-}MAG}$ | The delay between the AP and the MAG. | 2 ms [4] |
| $t_{AAA\text{-}Resp}$ | Delay due to AAA response message. | 1 ms |
| $t_{AAA\text{-}Req.}$ | Delay due to AAA request message. | 1 ms |
| $T_{AAA}$ | Delay due to AAA procedure. | 3 ms [4] |
| n, m | Number of hops between MAG-LMA and LMA-MAG respectively | 1-15 |
| $\beta_{RP}$ | LSR processing time. | 0.1 ms |
| $\beta_{MAG}$ | Processing time of MAG/LER router. | 0.2 ms |
| $\beta_{LMA}$ | Processing time of LMA/LER router. | 0.5 ms |
| $\alpha_{MAG}$ | Processing time of MAG router. | 0.2 ms [18] |
| $\alpha_{LMA}$ | Processing time of LMA router. | 0.5 ms [18] |
| $D_{Ul}$ | Upstream delay propagation in link l. | 2 ms |
| $D_{Dk}$ | Downstream delay propagation in link k. | 2 ms |
| $\lambda_{PR}$ | Send packet ratio | 170 packets/sec [19] |

### 4.2 Total Handover Delay

In this subsection we analyze the delay performance of the handover process for our PMIPv6/MPLS integration. The impact of handover on ongoing sessions is commonly characterized by handover delay, especially when we work with real time applications (e.g. Voice over IP, Video over Demand or IPTV) which are sensitive to packet delay and have important requirements of interruption time. For convenience, we define the parameters described in Table 8.

The general equation of the total handover delay in a Mobile IP protocols can be expressed as:

$$T_{HO} = T_{L2HO} + T_{MD} + T_{L3HO}. \qquad (1)$$

$T_{MD}$ is the interval from when an MN finishes Layer 2 handover to when it begins Layer 3 handover. In PM²PLS as in PMIPv6, as soon the MN is detected by the MAG with a L2 trigger, the L3 handover is initialized, so $T_{MD}$ can be considered zero.

$T_{L3HO}$ in PM²PLS when a bidirectional LSP exists between MAG and LMA can be expressed as:

$$T_{L3HO} = T_{AAA} + T_{REG} + T_{RA}, \qquad (2)$$

where the AAA process delay is as follows:

$$T_{AAA} = t_{AAA-Req.} + t_{AAA-Resp.} + \alpha_{AAA-Server}, \quad (3)$$

the binding update delay can be expressed as:

$$T_{REG} = t_{PBU} + t_{PBA} + \beta_{LMA} + \beta_{MAG} \quad (4)$$

where

$$t_{PBU} = t_{MAG,LMA} + (n)\,\beta_{RP} \quad (5)$$

$$t_{MAG,LMA} = \sum_{k=1}^{n} D_{Dk} \quad (6)$$

$$t_{PBA} = t_{LMA,MAG} + (m)\,\beta_{RP} \quad (7)$$

$$t_{LMA,MAG} = \sum_{l=1}^{m} D_{Ul} \quad (8)$$

finally,

$$T_{REG} = \sum_{k=1}^{n} D_{Dk} + \sum_{l=1}^{m} D_{Ul} + (n+m)\,\beta_{RP} + \beta_{LMA} + \beta_{MAG}. \quad (9)$$

When a bidirectional LSP is not established between MAG and LMA $T_{L3HO}$ can be calculated as follows:

$$T_{L3HO} = T_{AAA} + T_{REG} + T_{Bi-LSP-Setup} + T_{RA}, \quad (10)$$

where $T_{AAA}$ is the same as in (3), $T_{RA}$ is the same as in (16), and from (9) $T_{REG}$ can be expressed as:

$$T_{REG} = \sum_{k=1}^{n} D_{Dk} + \sum_{l=1}^{m} D_{Ul} + (n+m)\,\alpha_{RP} + \alpha_{LMA} + \alpha_{MAG}. \quad (11)$$

The latency introduced by LSP setup between the LMA and the MAG and vice versa ($T_{Bi-LSP-Setup}$) in PM²PLS can be expressed as the delay of one LSP setup, since the LMA initializes LSP setup between LMA and MAG after accepting PBU and sending PBA to the MAG (The LMA does not need to wait nothing else). When PBA arrives to the MAG, it initializes the LSP setup with LMA. We assume that when a LSP setup between MAG and LMA finishes, the LSP between LMA and MAG is already established, since it initialized before MAG to LMA LSP:

$$T_{Bi-LSP-Setup} = t_{RSVP-Resv} + t_{RSVP-Path} \quad (12)$$

where

$$t_{RSVP-Resv} = t_{MAG,LMA} + (n)\,\alpha_{RP}, \quad (13)$$

$$t_{RSVP-Path} = t_{LMA,MAG} + (m)\,\alpha_{RP}, \quad (14)$$

$t_{MAG,LMA}$ and $t_{LMA,MAG}$ are as in (6) and (8) respectively. Finally, $T_{BI-LSP-Setup}$ can be expressed as:

$$T_{Bi-LSP-Setup} = \sum_{k=1}^{n} D_{Dk} + \sum_{l=1}^{m} D_{Ul} + (n+m)\,\alpha_{RP}. \quad (15)$$

The delay by router advertisement message can be expressed as:

$$T_{RA} = t_{AP-MAG} + t_{WL}. \quad (16)$$

The L2 handover delay in an 802.11 WLAN access network can be expressed as:

$$T_{L2HO} = t_{Scanning} + t_{Autentication} + t_{Assocciation} \quad (17)$$

$T_{L3HO}$ in PMIPv6 is as in (2), with $T_{AAA}$ as in (3), $T_{REG}$ as in (11) and $T_{RA}$ as in (16). As mentioned above during a PMIPv6 handover is not executed neither Movement Detection (MD) nor Address Configuration (Included DAD).

### 4.3 Packet Loss During Handover

Packet Loss (PL) is defined as the sum of lost packets per MN during a handover. With (20) we can calculate the PL in a handover for a given MN.

$$PL_{PM^2PLS} = T_{PM^2PLS\,HO} * \lambda_{PR} \quad (20)$$

### 4.4 Operational Overhead

The operational overhead of PM²PLS is 4 bytes per packet (MPLS header size). PM²PLS reduces significantly the operational overhead with respect to PMIPv6 which has an operational overhead of 40 bytes when uses IPv4 or IPv6 in IPv6 encapsulation (over IPv6 Transport Network), 20 bytes of overhead when uses IPv4 or IPv6 in IPv4 encapsulation (over IPv4 Transport Network), 44 bytes when uses GRE tunnel over TN IPv6, or 24 bytes when uses GRE tunnel over IPv4 TN. A comparison of operational overhead between above schemes is summarized in Table 9.

Table 9: Operational Overhead

| Scheme and Tummeling Mechanism | Overhead per Packet | Description |
|---|---|---|
| PMIPv6 with IPv6 in IPv6 Tunnel | 40 | IPv6 header |
| PMIPv6 with IPv4 in IPv6 Tunnel | 40 | IPv6 header |
| PMIPv6 with IPv6 in IPv4 Tunnel | 20 | IPv4 header |
| PMIPv6 with IPv4 in IPv4 Tunnel | 20 | IPv4 header |
| PMIPv6 with GRE encapsulation (over TN IPv6) | 44 | IPv6 header + GRE header |
| PMIPv6 with GRE encapsulation (over TN IPv4) | 24 | IPv4 header + GRE header |
| PMIPv6/MPLS with VP Label (over TN IPv4 or IPv6) | 8 | 2 MPLS headers |
| PM²PLS (over TN IPv4 or IPv6) | 4 | MPLS headers |

### 4.5 Simulation Results

We compared PM²PLS, PMIPv6 [5] and PMIPv6/MPLS as proposed in [8]. We use typical values for parameters involved in above equations as shown in Table 8. Figure 6 shows the impact of hops between the MAG and the LMA in the handover delay. It can be observed that the handover delay increases with the number of hops. PMIPv6/MPLS is the scheme most affected by the number of hops because it integrates the LSP setup in encapsulated way and does not optimize this process. PMIPv6 and PM²PLS with a bidirectional LSP established between new MAP and LMA shown a comparable performance with slightly better response of PM$^2$PLS when the number of hops increase because binding update messages (i.e. PBU and PBA) are sent through bidirectional LSP established between the MAG and the LMA instead of using IP forwarding. Figure 7 shows the total packet loss during handover for above schemes. Since packet loss during handover is proportional to the handover latency, PM$^2$PLS also have the lowest packet loss ratio between compared schemes. For doing the packet loss simulation we consider a flow of VoIP [19].

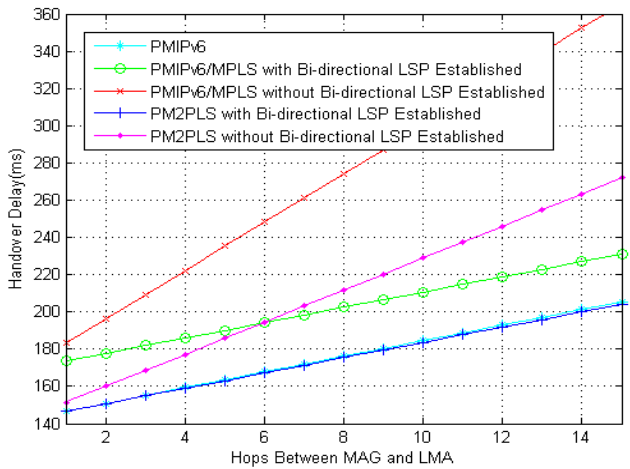

Fig. 7   802.11 handover process

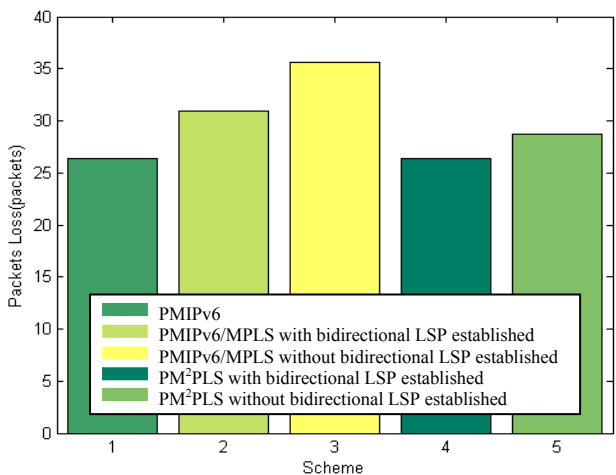

Fig. 8   Packet loss of PMIPv6, PMIPv6/MPLS, and PM2PLS during a handover.

## Conclusions

We proposed an integration of MPLS and PMIPv6 called PM$^2$PLS which optimizes the bidirectional LSP setup by integrating binding updates and bidirectional LSP setup in an optimized sequential way; we also used the LSP established between the MAG and the LMA for sending PBU and PBA messages when it exists. We compared the performance of PM$^2$PLS with single PMIPv6 and PMIPv6/MPLS as specified in [8]. We demonstrated that PM$^2$PLS has a lower handover delay than PMIPv6/MPLS, and slightly lower than the one of PMIPv6. The operational overhead in MPLS-based schemes is lower than single PMIPv6 schemes since uses LSPs instead of IP tunnelling. With MPLS integrated in a PMIPv6 domain, the access network can use intrinsic Quality of Service and Traffic Engineering capabilities of MPLS. It also allows the future use of DiffServ and/or IntServ in a PMIPv6/MPLS domain.

### Acknowledgments

This work was sponsored by the Colombian Institute of Science and Technology (COLCIENCIAS), http://www.colciencias.gov.co/ through the national program Young Researchers and Innovators "Virginia Gutiérrez de Pineda". We would like to thank MSc. Victor M. Quintero for his useful comments in the preliminary version of this paper published in the NTMS 2011.

**Carlos A. Astudillo** received his B.Sc. degree in Electronics and Telecommunications Engineering from the University of Cauca, Popayán, Colombia, in 2009. In 2010, he got a scholarship from the national program Young Researcher and Innovators "Virginia Gutiérrez de Pineda" of the Colombian Institute of Science and Technology - COLCIENCIAS. He is member of the New Technologies in Telecommunications R&D Group in the same University. Currently, he is a master student in Computer Science at State University of Campinas, Campinas, Brazil. His research interests are Mobility and Quality of Service in Wired/Wireless Networks and Policy-Based Network Management.

**Oscar J. Calderón** received his B.Sc. degree in Electronics and Telecommunications Engineering from the University of Cauca, Popayán, Colombia in 1996, He holds a specialist degree in Telematics Networks and Services (1999) and the Diploma of Advanced Studies (DEA) from the Polytechnic University of Catalonia (2005), Spain. He is full-professor and head of the Department of Telecommunications in the University of Cauca. He is member of the New Technologies in Telecommunications R&D Group in the same University. His research interests are Quality of Service in IP Networks, NGN.

**Jesús H. Ortiz** received his BSc. in Mathematical from the Santiago de Cali University, Colombia, Bsc. in Electrical Engineering from the University of Valle, Colombia and his PhD degree in Computer Engineering from the University of Castilla y la Mancha, Spain, in 1988, 1992 and 1998 respectively. Currently, he is assistant professor in the Universidad of Castilla y la Mancha, Spain in the area of Computer and Mobile Networks. He is reviewer and/or editor of several journals such as IAJIT, IJRRCS, IJCNIS, JSAT, and ELSEVIER.